\documentclass[twocolumn,groupaddress,aps,prxquantum,UTF8]{revtex4}
\usepackage{amsmath}
\usepackage{graphicx}
\usepackage{dcolumn}
\usepackage{color}
\usepackage{ulem}
\usepackage{bm}
\usepackage{braket}
\usepackage{array}
\usepackage[colorlinks,citecolor=blue]{hyperref}
\usepackage{multirow}
\usepackage{makecell}
\usepackage{booktabs}
\usepackage{graphicx}
\usepackage{multirow}
\renewcommand\thesection{\arabic{section}}
\renewcommand{\thesubsection}{\arabic{section}.\arabic{subsection}}
\usepackage{titlesec}
\titleformat{\section}
{\normalfont\normalsize\bfseries}{\thesection.}{1em}{}
\titleformat{\subsection}
{\normalfont\normalsize\itshape}{\thesubsection.}{1em}{}
\titleformat{\subsubsection}
{\normalfont\normalsize\itshape}{\thesubsubsection.}{1em}{}
\begin{document}

\title{Symmetry-Preserving Quadratic Lindbladian and Dissipation Driven Topological Transitions in Gaussian States}
\author{Liang Mao}
\affiliation{Institute for Advanced Study, Tsinghua University, Beijing,100084, China}
\author{Fan Yang}
\email{paullsc123@gmail.com}
\affiliation{Institute for Advanced Study, Tsinghua University, Beijing,100084, China}
\author{Hui Zhai}
\email{hzhai@tsinghua.edu.cn}
\affiliation{Institute for Advanced Study, Tsinghua University, Beijing,100084, China}
\date{\today}

\begin{abstract}

The dynamical evolution of an open quantum system can be governed by the Lindblad equation of the density matrix. In this paper, we propose to characterize the density matrix topology by the topological invariant of its modular Hamiltonian. Since the topological classification of such Hamiltonians depends on their symmetry classes, a primary issue we address is determining the requirement for the Lindbladian operators, under which the modular Hamiltonian can preserve its symmetry class during the dynamical evolution. We solve this problem for the fermionic Gaussian state and for the modular Hamiltonian being a quadratic operator of a set of fermionic operators. When these conditions are satisfied, along with a nontrivial topological classification of the symmetry class of the modular Hamiltonian, a topological transition can occur as time evolves. We present two examples of dissipation-driven topological transitions where the modular Hamiltonian lies in the AIII class with $U(1)$ symmetry and the DIII class without $U(1)$ symmetry. By a finite size scaling, we show that this density matrix topology transition occurs at a finite time. We also present the physical signature of this transition.

\end{abstract}

\maketitle

\section{Introduction}

In the past decades, topology has been extensively used to characterize the ground state wavefunction of a quantum Hamiltonian. In such situations, the topology of the wavefunction and the topology of the Hamiltonian are directly related. One of the most well-established situations is the insulators of free fermions \cite{Hasan10,TIreview,TIbook}. Another important lesson we have learned is the close relationship between the symmetry of a Hamiltonian and its topological classification \cite{Schnyder08,Kitaev09,Symmetry_RMP}. In many cases, topologically nontrivial states exist only when certain symmetries are enforced. For free fermion insulators and superconductors, this leads to the celebrated Altland-Zirnbauer ten-fold way classification \cite{tenfold} of topological states using time-reversal, particle-hole, and chiral symmetries \cite{Schnyder08,Kitaev09,Symmetry_RMP}.  

In recent years, there has been an increasing interest in studying topological properties for non-equilibrium dynamics. In some non-equilibrium situations, the quantum state under evolution remains a pure state when starting from a pure one, but is no longer an eigenstate of the system. A typical situation is quench dynamics \cite{Quench-1,Quench-2,Quench-3,Quench-4,Quench-5,Quench-6,Quench-7,Quench-8}, where an initial pure state undergoes unitary dynamical evolution governed by the system's Hamiltonian. It has been shown that a properly defined topology of the wavefunction dynamics can still reveal the topology of the system's Hamiltonian \cite{Quench-1,Quench-2,Quench-3,Quench-4,Quench-5,Quench-6,Quench-7,Quench-8}. Nevertheless, in more generic non-equilibrium situations, the quantum state is not even a pure state but a mixed state described by a density matrix. These situations include finite temperature and open systems with dissipation.   

An open quantum system is described by a density matrix $\hat{\rho}$ whose evolution is governed by the Lindblad equation \cite{lindblad1,lindblad2}. Since the Lindbladian evolution can be viewed as evolution under non-hermitian Hamiltonian in doubled Hilbert space, there are works on open system topology by considering the topology of Lindbladian itself \cite{top_lind1,top_lind2,top_lind3}. At sufficiently long time, an open system can reach a non-equilibrium steady state that does not evolve. There are also extensive works focusing on the topological properties of the non-equilibrium steady states \cite{top_dis1,top_dis2,top_dis3,top_dis4,top_dis5,top_dis6,top_dis7,top_dis8,top_dis9}.

On the other hand, the topology of density matrix itself attracts attention \cite{top_dens1,top_dens2,top_dens3,top_dens4,top_dens5,top_dens6,top_dens7,top_dens8}. The basic idea is to consider the modular Hamiltonian $\hat{K}$ by writing $\hat\rho=e^{-\hat{K}}$. The modular Hamiltonian is always a Hermitian operator, and the topology of the modular Hamiltonian can characterize the topology of the density matrix \cite{top_dens3,top_dens5,top_dens8}. The modular Hamiltonian changes in time when the density matrix evolves under the Lindbladian dynamics. In this work, we address the issue of \textit{whether the modular Hamiltonian can undergo a topological transition as time evolves in the Lindbladian dynamics of an open system}. This question concerns the entire dissipation-driven dynamical process instead of the long-time steady state. The answer to this question should depend on both the Lindbladian operator and the choice of the initial state. This distinguishes our work from the previous studies of open-system topology \cite{top_lind1,top_lind2,top_lind3,top_dis1,top_dis2,top_dis3,top_dis4,top_dis5,top_dis6,top_dis7,top_dis8,top_dis9}.

Here we consider the Gaussian state of a set of fermion operators. A gaussian state can be viewed as a non-equilibrium generalization of free-fermion state, whose modular Hamiltonian is a quadratic operator of a set of fermion operators \cite{gauss1,gauss2}. Hence, we can utilize the existing knowledge of classifying such quadratic fermionic Hamiltonian.  It is known that the topological classification of such quadratic Hamiltonians requires understanding its symmetry class \cite{Symmetry_RMP,Schnyder08,Kitaev09}. In this work, we consider a generic non-equilibrium situation where the modular Hamiltonian of the initial state have no relation with the system's Hamiltonian and can lie in a different symmetry class. Hence, in order to address the dissipation-driven topological transition, we have to first address another issue: \textit{under what conditions the Lindbladian evolution can preserve the symmetry of a modular Hamiltonian}?

\section{Symmetry preserving Lindbladian}\label{2}

The time evolution of the density matrix of an open system is governed by the Lindblad equation \cite{lindblad1,lindblad2}
\begin{equation}
\frac{d\hat{\rho}}{dt}=-i[\hat{H},\hat{\rho}]+\sum\limits_{\mu}(2\hat{L}_\mu\hat{\rho}\hat{L}^\dag_\mu-\{\hat{L}^\dag_\mu\hat{L}_\mu,\hat{\rho}\}). 
\end{equation}
Here $\hat{H}$ refers to the system Hamiltonian and $\hat{L}_\mu$s are the dissipation operators. We consider Gaussian initial state $\hat\rho(t=0)$ of a set of fermion operators $\{\hat{c}^\dag_i\}$. The Hamiltonian $\hat{H}$ is a quadratic operator of these fermion operators, and all $\hat{L}_\mu$ are linear in these fermion operators, under which a Gaussian state can remain a Gaussian one during the Lindblad time evolution \cite{gauss2,correlation1,correlation2}. In this work, we deal with the most generic situations that $\hat{K}$, $\hat{H}$ and $\hat{L}_\mu$ do not commute with each other. 

We first study the cases that both the modular Hamiltonian and the Lindbladian possess a global $U(1)$ symmetry. This $U(1)$ symmetry can be either a spin $U(1)$ or a charge $U(1)$ symmetry. When interpreted as a charge (spin) $U(1)$ symmetry, the corresponding class describes insulators (superconductors).  For superconductors with the spin $U(1)$ symmetry, one can always apply a particle-hole transformation to map the spin $U(1)$ symmetry into a charge $U(1)$ symmetry. Therefore, we present the following discussions in the context of the charge $U(1)$ symmetry.

With the presence of the charge $U(1)$ symmetry, we can write $\hat{K}=\sum_{ij}K_{ij}\hat{c}^\dag_i\hat{c}_j$, and $\hat{H}=\sum_{ij}h_{ij}\hat{c}^\dag_i\hat{c}_j$. Each dissipation operator should be either pure loss term as a superposition of annihilation operators, denoted by $\hat{L}^l_\mu=\sum_{\mu,i}D^{l}_{\mu i}\hat{c}_i$, or pure gain term as a superposition of creation operators  $\hat{L}^g_\mu=\sum_{\mu,i}D^{g}_{\mu i}\hat{c}^\dag_i$. In other words, each single dissipation operator cannot contain both creation and annihilation operators. $K_{ij}$, $h_{ij}$, $D^{l}_{\mu i}$ and $D^{g}_{\mu i}$ are matrices or vectors written in the single particle bases.  

Below we should study how the modular Hamiltonian matrix $K$ evolves in time. Nevertheless, the matrix $K$ obeys a non-linear equation which complicates the analyzing process. On the other hand, another important property of a Gaussian state is that the two-point correlation contains all the information of this state, and all the higher-order correlations can be expressed in terms of two-point corrections. Thus, instead of considering the dynamics of the matrix $K_{ij}$, we can consider the correlation matrix $C$, defined as $C_{ij}=\mathrm{Tr}(\hat{c}^\dag_i\hat{c}_j\hat{\rho})$. It is important to note that the correlation matrix $C$ obeys a linear equation. For cases with the $U(1)$ symmetry, the correlation matrix and the modular Hamiltonian matrix are related by \cite{correlation1,correlation2}
\begin{equation}
C=\frac{1}{e^{K^\text{T}}+1}, \label{CK}
\end{equation}
where ``$\text{T}$" stands for transpose. Following the Lindblad equation, it can be shown that the correlation matrix obeys the following linear equation \cite{correlation1,correlation2,correlation3}
\begin{equation}
\frac{dC}{dt}=XC+CX^\dag+2M^g
\end{equation}
where the matrix $X$ is defined as $X=i H^\text{T}-(M^{l})^\text{T}-M^g$ and $H$ is the physical Hamiltonian matrix. Here the matrix $M^l$ is from the loss terms, defined as $M^l_{ij}=\sum_\mu D^{l*}_{\mu i}D^{l}_{\mu j}$, and the matrix $M^g$ is from the gain terms  $M^g_{ij}=\sum_\mu D^{g*}_{\mu i}D^{g}_{\mu j}$.

Following the ten-fold way classification, we consider the time-reversal symmetry $\mathcal{T}$, the particle-hole symmetry $\mathcal{C}$ and the chiral symmetry $\mathcal{S}$ of the modular Hamiltonian matrix $K$. Here we should clarify that the time-reversal symmetry should not be understood as a physical time reversal of the system. Instead, it merely means an antiunitary transformation that changes $K$ by $U^\dag_\mathcal{T} KU_\mathcal{T}=K^*$, where $U_{\mathcal{T}}$ is the unitary part of $\mathcal{T}$ symmetry. 

First we derive the conditions to preserve $\mathcal{T}$ symmetry. Using Eq. \ref{CK}, $\mathcal{T}$ symmetry leads to $C^\text{T}U_\mathcal{T}=U_\mathcal{T}C$. That is to say, initially, $C^\text{T}U_\mathcal{T}-U_\mathcal{T}C=0$. If the Lindblad evolution can keep the $\mathcal{T}$ symmetry of the modular Hamiltonian, it requires $d(C^\text{T}U_\mathcal{T}-U_\mathcal{T}C)/dt=0$. It is easy to show that 
\begin{align}
&\frac{d}{dt}(C^\text{T}U_\mathcal{T}-U_\mathcal{T}C)=(X^*U_\mathcal{T}-U_\mathcal{T}X)C\nonumber\\
&+C^\text{T}(X^\text{T}U_\mathcal{T}-U_\mathcal{T}X^\dag)+2(M^gU_\mathcal{T}-U_\mathcal{T}M^g).
\end{align}
Since we consider generic initial states with $\mathcal{T}$ symmetry, the requirement $d(C^\text{T}U_\mathcal{T}-U_\mathcal{T}C^\text{T})/dt=0$ leads to a set of sufficient conditions 
$X^*U_\mathcal{T}-U_\mathcal{T}X=0$, $X^\text{T}U_\mathcal{T}-U_\mathcal{T}X^\dag=0$ and $M^gU_\mathcal{T}-U_\mathcal{T}M^g=0$. These conditions can be further simplified as 
\begin{align}
U^\dag_\mathcal{T} H U_\mathcal{T}=-H^*, U^\dag_\mathcal{T}  M^l U_\mathcal{T} =(M^{l})^*, U^\dag_\mathcal{T}  (M^{g})^*U_\mathcal{T} =M^g.  \label{condition-T}
\end{align}
In other words, if the modular Hamiltonian of the initial state has $\mathcal{T}$ symmetry, and the Hamiltonian and the dissipation operators satisfy Eq. \ref{condition-T}, the $\mathcal{T}$ symmetry will be preserved during the entire Lindbladian dynamics. Especially, we note that the required symmetry property for the physical Hamiltonian $H$ is different compared with the symmetry property of the modular Hamiltonian $K$. 

Next, we consider the particle-hole symmetry $\mathcal{C}$. With this symmetry the modular Hamiltonian matrix transfers as $U^\dag_\mathcal{C} K U_\mathcal{C}=-K^*$ under a unitary matrix $U_\mathcal{C}$. This is equivalent to $C^\text{T}U_\mathcal{C}+U_\mathcal{C}C-U_\mathcal{C}=0$. Hence, in order to preserve the particle-hole symmetry, we require $d(C^\text{T}U_\mathcal{C}+U_\mathcal{C}C-U_\mathcal{C})/dt=0$. Similar analysis as above leads to following conditions 
\begin{align}
U^\dag_\mathcal{C} H U_\mathcal{C}=-H^*, U^\dag_\mathcal{C} M^l U_\mathcal{C}=M^{g}, U^\dag_\mathcal{C} (M^{g})^*U_\mathcal{C}=(M^l)^*.  \label{condition-C}
\end{align}
We note that preserving the particle-hole symmetry requires simultaneously presence both the loss and the gain terms. 

Finally we consider the chiral symmetry $\mathcal{S}$ under which the modular Hamiltonian matrix transfers as $U^\dag_\mathcal{S} K U_\mathcal{S}=-K$, equivalent to $C^\text{T}U_\mathcal{S}+U_\mathcal{S}C^\text{T}-U=0$. Hence, to preserve the chiral symmetry, we require $d(C^\text{T}U_\mathcal{S}+U_\mathcal{S}C^\text{T}-U_\mathcal{S})/dt=0$. Following the same spirit, we arrive at
\begin{align}
U^\dag_\mathcal{S} H U_\mathcal{S}=H, U^\dag_\mathcal{S} M^l U_\mathcal{S}=(M^{g})^*, U^\dag_\mathcal{S} (M^{g})^*U_\mathcal{S}=M^l.  \label{condition-S}
\end{align}
Time-reversal and the particle-hole symmetries automatically guarantees the chiral symmetry by taking $U_\mathcal{S}=U_\mathcal{C}U_\mathcal{T}^*$. As a self-consistent check, it is easy to prove that Eq. \ref{condition-T} and Eq. \ref{condition-C} automatically ensure Eq. \ref{condition-S}.

Eq. \ref{condition-T}, Eq. \ref{condition-C} and Eq. \ref{condition-S} respectively give the conditions for preserving $\mathcal{T}$, $\mathcal{C}$ and $\mathcal{S}$ symmetries of the modular Hamiltonian when the system and initial states are $U(1)$ symmetric. 
Next, we move to the situation without charge or spin $U(1)$ symmetry, such as the cases with fermion pairing between same spins. In this case, we use the Nambu spinor by introducing $\hat{\Psi}=(\hat{c}_1,\dots,\hat{c}_N,\hat{c}^\dag_1,\dots,\hat{c}^\dag_N)^\text{T}$, and we write the physical and the modular Hamiltonians into the Bogoliubov form as $\hat{H}=\hat{\Psi}^\dag H\hat{\Psi}$ and $\hat{K}=\hat{\Psi}^\dag K\hat{\Psi}$. Unlike the cases with $U(1)$ symmetry, the dissipation operators can be a superposition of both loss and gain as $\hat{L}_\mu=D^l_{\mu i}\hat{c}_i+D^g_{\mu i}\hat{c}^\dag_i$. We introduce the correlation matrix as $C$ as $C_{ij}=\mathrm{Tr}( \hat{\Psi}^\dag_i\hat{\Psi}_j\hat{\rho})$ that includes anomalous correlations. Now the correlation matrix and the modular Hamiltonian matrix are related by
\begin{equation}
C=\frac{1}{e^{2 K^\text{T}}+1}. \label{CK2}
\end{equation}
Similarly, we can define the matrix $M^l$ and $M^g$ as introduced above and two extra matrices $M^{lg}_{ij}=\sum_\mu D^{l*}_{\mu i}D^{g}_{\mu j}$ and $M^{gl}_{ij}=\sum_\mu D^{g*}_{\mu i}D^{l}_{\mu j}$. Following the Lindblad equation, the correlation matrix $C$ now obeys the linear equation 
\begin{equation}
\frac{dC}{dt}=XC+CX^\dag+2W.
\end{equation}
Here the matrix $X=2iH^\text{T}-M^\text{T}-W$, and the matrices $M$ and $W$ are respectively written 
\begin{equation}
M=\left(\begin{array}{cc}M^l & M^{lg} \\ M^{gl} & M^g\end{array}\right), \   \ W=\left(\begin{array}{cc}M^g & M^{gl} \\ M^{lg} & M^l\end{array}\right).
\end{equation} 

Since the Bogoliubov Hamiltonian automatically possesses the particle-hole symmetry, we only need to investigate the time-reversal symmetry. The derivation of the symmetry preserving condition is very similar to the case with $U(1)$ symmetry above. The results are
\begin{align}
U^\dag_\mathcal{T} H U_\mathcal{T}=-H^*, U^\dag_\mathcal{T}  M U_\mathcal{T} =M^*, U^\dag_\mathcal{T}  W^*U_\mathcal{T} =W.  \label{condition-T-no-U1}
\end{align}
Hence we have succeeded in deriving the conditions for Lindbladian to preserve $\mathcal{T}$, $\mathcal{C}$ and $\mathcal{S}$ symmetries with and without $U(1)$ symmetry.

\section{Dissipation-Driven Density Matrix Topological transition}\label{3}

Now we give concrete examples of topological transition. It is easy to see that when the following three conditions are satisfied, the modular Hamiltonian can undergo a topological transition during the dissipation dynamics.
\begin{itemize}
	\item The initial modular Hamiltonian must lie in a symmetry class and dimension which hosts nontrivial topological classification
	\item The Lindbladian must satisfy the aformentioned conditions to preserve this symmetry class
	\item The modular Hamiltonians of the initial state and the long-time steady state are both gapped and have different topological numbers
\end{itemize}
  Below we will discuss two examples.

\subsection{An Example of 1D, AIII Class}
 Our first example is a one-dimensional model in the AIII class, which is the celebrated Su-Schrieffer-Heeger (SSH) model \cite{SSH} with dissipations. This one-dimensional lattice contains two sites in each unit cell, denoted by site-$A$ and -$B$, and the modular Hamiltonian takes the form of the SSH model as
\begin{equation}\label{ssh}
\hat{K}(t=0)=\sum\limits_{i}J_1\hat{c}^\dag_{i,A}\hat{c}_{i,B}+J_2\hat{c}^\dag_{i,B}\hat{c}_{i+1,A}+\text{h.c.}.
\end{equation}
This class possesses the chiral symmetry and the SSH model has the charge $U(1)$ symmetry. Hence, the physical Hamiltonian and the dissipation operators have to satisfy Eq. \ref{condition-S}.
Here we choose the set of fermion operator basis as $\{\dots,\hat{c}_{i,A},\hat{c}_{iB},\hat{c}_{i+1,A},\hat{c}_{i+1,B},\dots\}$. Under this basis, the $K$ matrix and the corresponding $U_\mathcal{S}$ matrix are written as
\begin{equation}
K = 	\begin{pmatrix}
		&J_1&&&\\
		J_1&&J_2&&\\
		&J_2&&J_1&\\
		&&J_1&&\\
		&&&&\ddots
	\end{pmatrix}; 
	U_\mathcal{S} = 
	\begin{pmatrix}
		1&&&&\\
		&-1&&&\\
		&&1&&\\
		&&&-1&\\
		&&&&\ddots
	\end{pmatrix} \nonumber
\end{equation}
It is easy to see that this modular Hamiltonian obeys the chiral symmetry condition $U^\dag_\mathcal{S}KU_\mathcal{S}=-K$.

\begin{figure}
    \centering
    \includegraphics[width=0.45\textwidth]{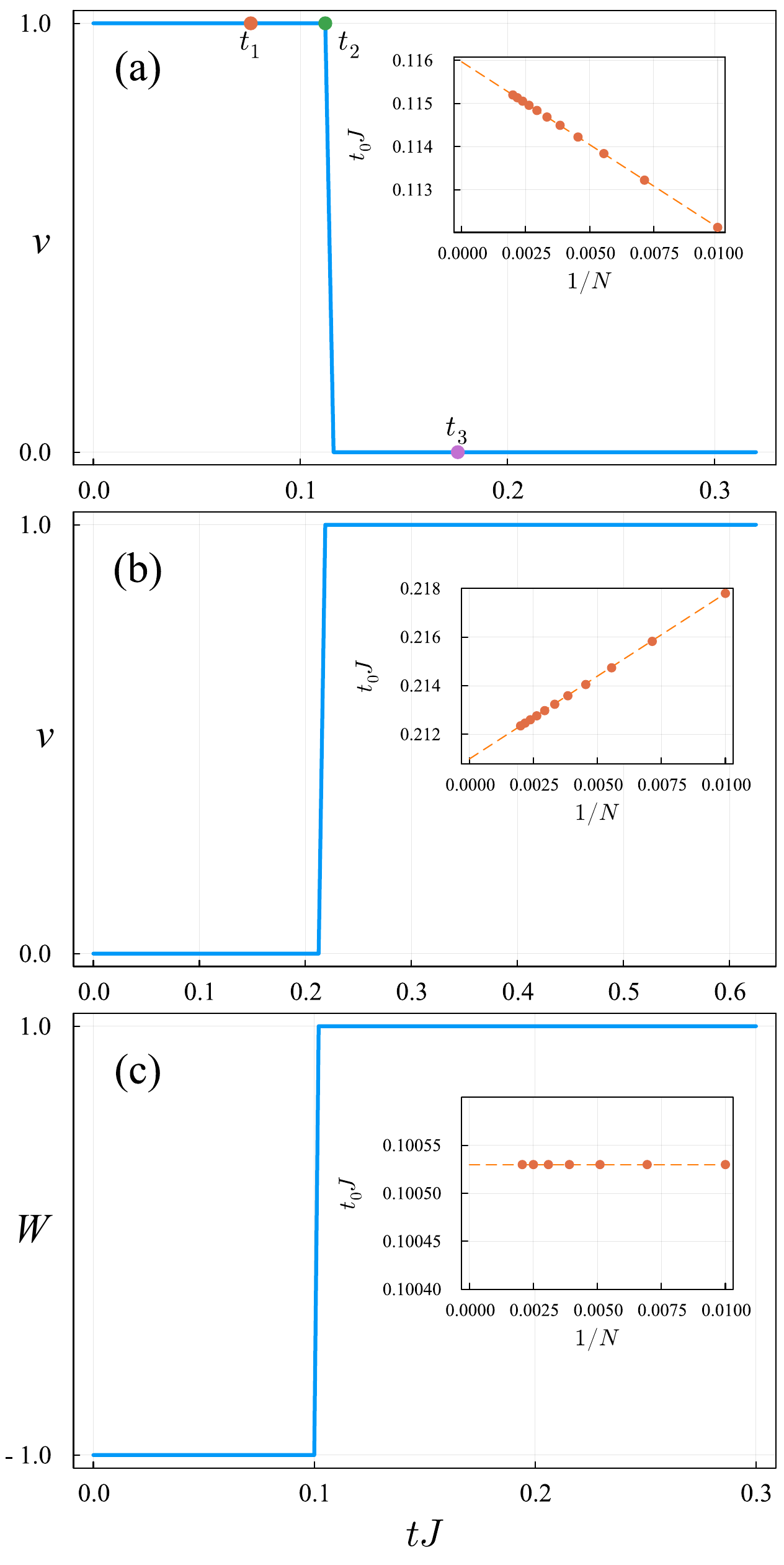}
    \caption{Topological transition of the density matrix during the Lindbladian evolution. The vertical axis is the topological invariant of the modular Hamiltonian and the horizontal axis is time $t$ in unit of hopping $J$.  (a-b)  Example I. The modular Hamiltonian is a one-dimension SSH model in the AIII class. The topological invariant is the winding number $v$ defined in Eq.\ref{winding}. $J=(J_1+J_2)/2$. (a) $J_1=0.8J$ and $J_2=1.2J$, $\gamma_1=0.8J$ and $\gamma_2=0.4J$; (b) $J_1=1.2J$ and $J_2=0.8J$, $\gamma_1=0.4J$ and $\gamma_2=0.8 J$. $\delta=0.1J$ for both (a) and (b) and $N=100$ unit cells is taken for simulation. (c) Example II. The modular Hamiltonian belongs to two-dimensional square lattice pairing model in DIII class. The topological invariant is the Fu-Kane invariant $W$ defined in Eq.\ref{fk}.  We set $\mu=1.0J$, $\Delta_0=1.0J$, $w=0.5J$, $\gamma_1=0.4J$ and $\gamma_2=0.6J$. $N=10\times 10$ unit cells is taken for simulation. The insets show the transition time $t_0J$ as a function of $1/N$, where $N$ is the system size of numerical simulation. This is to extract the transition time at $N\rightarrow\infty$.  The states at $t_1$, $t_2$ and $t_3$ are taken to simulate the different behavior of spectrum property and $\widetilde{S}_n$, as shown in Fig. \ref{entropy}.}
     \label{transition}
\end{figure}

We consider a simple physical Hamiltonian $\hat{H}=\delta\sum_{i}(\hat{c}^\dag_{i,A}\hat{c}_{i,A}-\hat{c}^\dag_{i,B}\hat{c}_{i,B})$. The $H$-matrix is a diagonal matrix denoted by $H=\text{diag}(\delta,-\delta,\delta,-\delta,\dots)$ and it satisfies $U^\dag_\mathcal{S}H U_\mathcal{S}=H$. We choose two types of loss operators $\hat{L}^l_i=\sqrt{\gamma_1}(\hat{c}_{i,A}+\hat{c}_{i,B})$ and $\hat{L}^l_i=\sqrt{\gamma_2}(\hat{c}_{i,B}+\hat{c}_{i+1,A})$, and two types of gain operators $\hat{L}^g_i=\sqrt{\gamma_1}(\hat{c}^\dag_{i,A}-\hat{c}^\dag_{i,B})$, $\hat{L}^g_i=\sqrt{\gamma_2}(\hat{c}^\dag_{i,B}-\hat{c}^\dag_{i+1,A})$. Hence, the two dissipation matrices $M^l$ and $M^g$ are respectively given by
\begin{align}
&M^l =
	\begin{pmatrix}
		\gamma_1+\gamma_2&\gamma_1&&&\\
		\gamma_1&\gamma_1+\gamma_2&\gamma_2&&\\
		&\gamma_2&\gamma_1+\gamma_2&\gamma_1&\\
		&&&&\ddots\\
	\end{pmatrix}; \nonumber\\
&M^g =
\begin{pmatrix}
	\gamma_1+\gamma_2&-\gamma_1&&&\\
	-\gamma_1&\gamma_1+\gamma_2&-\gamma_2&&\\
	&-\gamma_2&\gamma_1+\gamma_2&-\gamma_1&\\
	&&&&\ddots\\
\end{pmatrix}. \nonumber
\end{align}
It is easy to verify that these two matrices satisfy $U^\dag_\mathcal{S} M^l U_\mathcal{S}=(M^{g})^*$ and  $U^\dag_\mathcal{S} (M^{g})^*U_\mathcal{S}=M^l$. Hence, we show that this physical Hamiltonian and the dissipation operators can keep the chiral symmetry of the modular Hamiltonian during the evolution. 

Moreover, for the Lindbladian we considered, we find that, when $\gamma_1>\gamma_2$, the modular Hamiltonian of the steady state is topologically trivial, and when $\gamma_2>\gamma_1$, the modular Hamiltonian of the steady state is topologically nontrivial. Hence, if we choose the initial modular Hamiltonian as a topologically nontrivial one for the former case and a topological trivial one for the latter case, a dissipation dynamics driven topological transition should occur, as shown in Fig.  \ref{transition}(a) and (b). In these figures, we plot the topological winding number $v$ of the modular Hamiltonian as time evolves.
To define the topological winding number, we first define matrix $Q(k) = 1-2\sum_{\alpha}\ket{\psi_k^\alpha}\bra{\psi_k^\alpha}$, where $\ket{\psi_k^\alpha}$s are eigenstates of modular Hamiltonian matrix $K$ and the sum runs over bands with negative energy.
In the chiral symmetric case, $Q(k)$ can be transformed to block off-diagonal 
\begin{align}
	Q(k) = 
	\begin{pmatrix}
		0&q(k)\\
		q^\dagger(k)&0
	\end{pmatrix}.
\label{Q}
\end{align}
Then the topological winding number is defined as
\begin{align}
	v = \frac{i}{2\pi}\int_{\text{BZ}}\mathrm{d}k
	\mathrm{Tr}(q^{-1}\partial_kq).
	\label{winding}
\end{align}
The integration runs over first Brillounin zone. According to the numerical results, a jump of the winding number can be found in the intermediate time \cite{code}. We note that in the case, the topological invariant of the modular Hamiltonian is directly measurable through the ensemble geometric phase \cite{top_dens3}.

\subsection{An Example of 2D, DIII Class}
 This example considers a two-dimensional model in the DIII class. Here we consider spin-$1/2$ fermions in two-dimensional square lattice, with $i=(i_x,i_y)$ labeling each site. The initial modular Hamiltonian is chosen as follows: 
\begin{align}
\hat{K}=&-J\sum\limits_{\langle ij\rangle\sigma}\hat{c}^\dag_{i\sigma}\hat{c}_{j\sigma}-\mu\sum\limits_{i}\hat{c}^\dag_{i\sigma}\hat{c}_{i\sigma}\nonumber\\
&+\sum\limits_{\langle ij\rangle\sigma}\Delta_{\langle ij\rangle\sigma}\hat{c}_{i\sigma}\hat{c}_{j\sigma}+\Delta^*_{\langle ij\rangle\sigma}\hat{c}^\dag_{j\sigma}\hat{c}^\dag_{i\sigma}.
\end{align}
Here $\langle ij\rangle$ denotes pairs of two nearest neighboring sites. For $\sigma=\uparrow$, we have $\Delta_{\langle ij\rangle\uparrow}=\pm \Delta_0$ if $j_x=i_x\pm 1$ and $j_y=i_y$ and $\Delta_{\langle ij\rangle\uparrow}=\pm i\Delta_0$ if $j_y=i_y\pm 1$ and $j_x=i_x$. This gives a $p_x+ip_y$ pairing for spin-$\uparrow$ between two neighboring sites. Similarly, we introduce a $p_x-ip_y$ pairing for spin-$\downarrow$ between neighboring sites. The topological invariant of this Hamiltonian is given by the Fu-Kane invariant \cite{f-k1,f-k2,f-k3,f-k4,f-k5}, protected by the time-reversal symmetry $\mathcal{T}$. With the presence of time-reversal symmetry and particle-hole constraint, the $Q(k)$ matrix can also be transformed block off-diagonal as Eq.\ref{Q}. Then the Fu-Kane invariant is defined as
\begin{align}
	W = \prod_\Gamma\frac{\mathrm{Pf}[q(\Gamma)]}{\sqrt{\det[q(\Gamma)]}}.
	\label{fk}
\end{align}
Here $\mathrm{Pf}[\cdots]$ is the Pfaffian and $\Gamma$ runs over time-reversal symmetric points $(0,0),(\pi,0),(0,\pi)$ and $(\pi,\pi)$.
The topological invariant $W$ is a $Z_2$ index, where $+1$ and $-1$ stand for topologically trivial and nontrivial cases, respectively. Moreover, it is easy to see that this model has neither the charge $U(1)$ nor the spin $U(1)$ symmetries. 

Here we consider the physical Hamiltonian and the dissipation operator as follows
\begin{align}
&\hat{H}=w\sum\limits_{\langle ij\rangle}(-1)^\sigma \hat{c}^\dag_{i\sigma}\hat{c}_{j\sigma},\\
&\hat{L}_i=\sqrt{\gamma_1}\hat{c}_{i\sigma}+(-1)^\sigma\sqrt{\gamma_2}(\hat{c}^\dag_{i\sigma}+\alpha_{ij}\hat{c}^\dag_{j\sigma}),
\end{align}
where $(-1)^\sigma=1$ for $\sigma=\uparrow$ and $(-1)^\sigma=-1$ for $\sigma=\downarrow$. In the definition of $\hat{L}_i$, $\alpha_{ij}=1$ if $j_x=i_x+1$ and $j_y=i_y$, and $\alpha_{ij}=(-1)^\sigma \mathrm{i}$ if $j_y=i_y+1$ and $j_x=i_x$, and $\alpha_{ij}=0$ otherwise. It can be shown that this choice of the Lindblad operator satisfies the condition Eq. \ref{condition-T-no-U1} for preserving the time-reversal symmetry in this model. The steady state of this Lindblad operator is topologically trivial. Hence, when the modular Hamiltonian of the initial state is topologically nontrivial, a transition must occur in the intermediate time, as shown in Fig. \ref{transition}(c).

\subsection{Transition Time}

In the above examples, we observe dissipation driven topological transitions at finite time. However, these results seem to contradict to its closed system counterpart. In a closed system, symmetry-preserving local unitary evolution cannot drive a topological transition at finite time. Then, a natural question arises: in the thermodynamic limit whether the topological transition really happen at finite time or it takes place at $t\rightarrow \infty$. 

To answer this question, we numerically calculate the transition time of both examples for various system size and then extrapolate the transition time to infinite system. The insets of Fig. \ref{transition} shows the transition time $t_0J$ as function of $1/N$. In both examples, we find that the transition time linearly depends on $1/N$ and saturates to a finite value as $N\rightarrow \infty$. Therefore we can conclude that the density matrix topological transition can take place at finite time, a situation opposite to the close system analogy. This is understandable because under the Lindbladian evolution, the modular Hamiltonian does not undergo a unitary transformation and the change of its topology is not forbidden.  Similar issue has also discussed in recent papers on symmetry protected topological phases of mixed state by using different metric \cite{aspt1,aspt2}.

\subsection{Physical Consequence of the Transition}\label{4}
\begin{figure}[!t]
	\centering
	\includegraphics[width=0.48\textwidth]{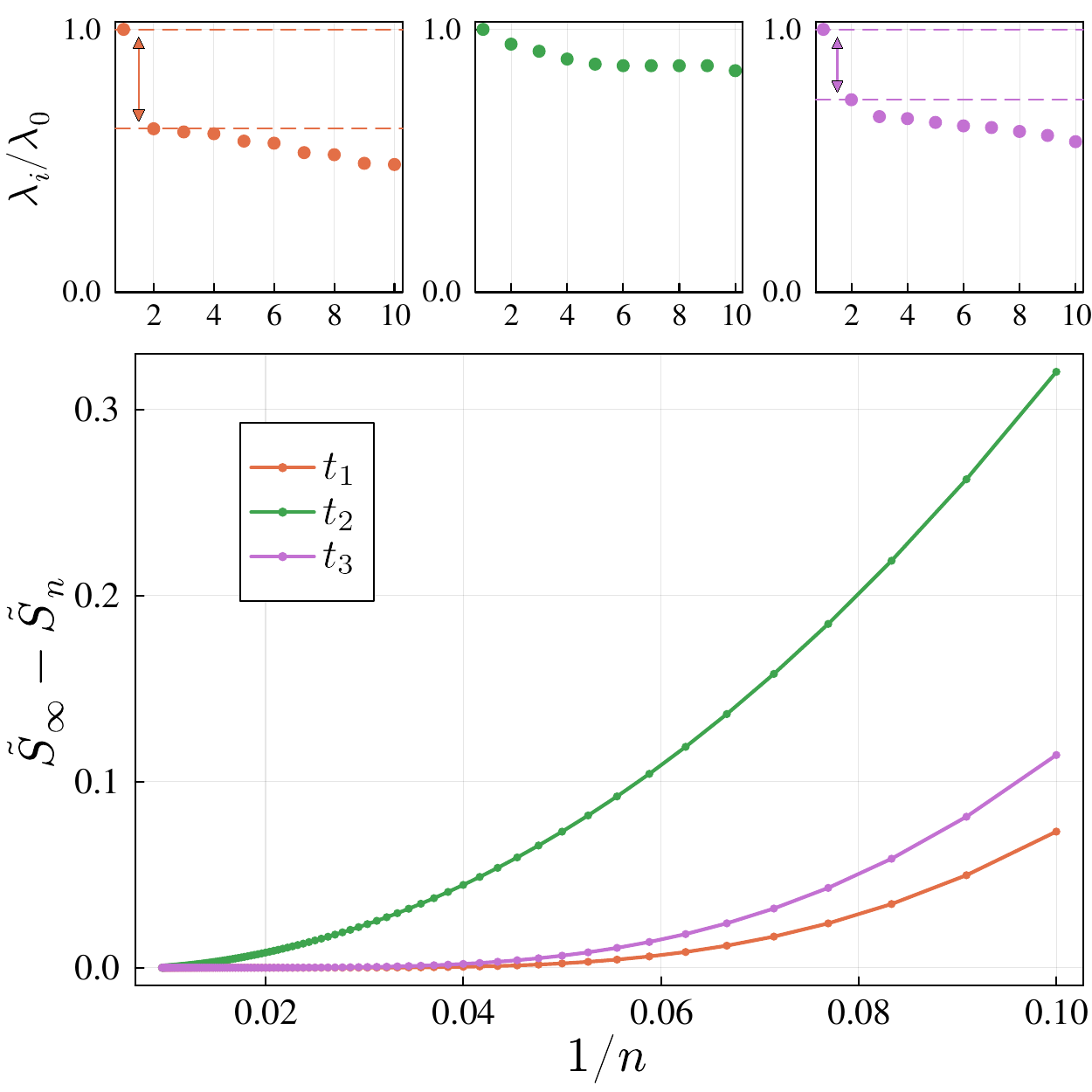}
	\caption{Top: The eigenvalues of density matrix at each time denoted in Fig \ref{transition} in descending order. Here $\lambda_i$ labels the eigenvalue, where $\lambda_0$ is the largest one. Bottom: Large $n$ behavior of $\widetilde{S}_\infty-\widetilde{S}_n$ at these three different time, where we have chosen $\widetilde{S}_\infty=\widetilde{S}_{200}$ for convenience. $t_2$ is the transition time. Here we consider the closed boundary condition.  }
	\label{entropy}
\end{figure}

Now we discuss physical consequence of this topological transition. The top three figures of Fig \ref{entropy} shows the eigenvalues of the density matrix at three different time slots denoted in Fig. \ref{transition}(a). These plots show that when the modular Hamiltonian is gapped, all other eigenvalues are separated from $\lambda_0$ by a finite gap, where $\lambda_0$ is the largest eigenvalue of the density matrix. This gap is also called the purity gap\cite{top_dis1,top_dis9,top_dens3}. Nevertheless, at the transition time, the modular Hamiltonian becomes gapless, and therefore, the purity gap vanishes in the thermodynamic limit.

To reveal the difference between the presence and the absence of the purity gap, we define the \textit{modified R\'enyi entropy} as
\begin{align}
	\widetilde{S}_n = -\frac{1}{n}\log(\mathrm{Tr}(\hat{\rho}^n)).  \label{modifiedRE}
\end{align} 
It is easy to show that when $n\rightarrow \infty$, $S_\infty=-\log\lambda_0$. For a density matrix with finite purity gap $\Delta$, $\widetilde{S}_n$ approaches $\widetilde{S}_\infty$ exponentially fast as
\begin{align}
	\widetilde{S}_n-\widetilde{S}_\infty &= -\frac{1}{n}\log(1+(\lambda_1/\lambda_0)^n+\cdots)\notag\\
	&\sim\mathcal{O}(e^{-n\Delta}/n). \label{Sngap}
\end{align}
However, the behavior is different in absence of the purity gap. For simplicity, let us consider the ground state of the modular Hamiltonian being $\epsilon$-fold degenerate, then
\begin{align}
	\widetilde{S}_n &=-\frac{1}{n}\log(\epsilon \lambda_0^n+\lambda_1^n+\cdots)\notag\\
	&=\widetilde{S}_\infty -\frac{1}{n}\log\epsilon+\mathcal{O}(e^{-\Delta n}/n). 
\end{align}
Here the leading order is $\mathcal{O}(1/n)$ instead of $\mathcal{O}(e^{-\Delta n}/n)$. So $\widetilde{S}_n$ converges much slower than the cases with gapped unique ground state. The gapless cases have similar behavior of $1/n$ convergence. We numerically verify the large $n$ behavior of $\widetilde{S}_n$ in Fig. \ref{entropy} for three time slots whose density matrix spectrums are displayed above. It shows that the convergence to $\widetilde{S}_\infty$ is much slower at the transition time where density matrix spectrum becomes gapless. This manifests the dissipation driven topological transition. 

Here we remark that it is important to define the modified R\'enyi entropy as Eq. \ref{modifiedRE} using prefactor $1/n$ instead of $1/(n-1)$ in the conventional definition of R\'enyi entropy. Because if we use the conventional definition, the gapped situation Eq. \ref{Sngap} will also show a $1/n$ term such that it cannot be distinguished from the gapless situation. We also note that, for open boundary condition, the topological modular Hamiltonian possesses gapless edge modes. Therefore, $\widetilde{S}_n$ also displays the $1/n$ convergence to $\widetilde{S}_\infty$, where the slop of $1/n$ term reveals the number of gapless edge modes.

\section{Outlook}\label{5}

In summary, we have revealed a novel phenomenon of dissipation dynamics driven transition of the density matrix topology, characterized by the topological invariant of the modular Hamiltonian of a Gaussian state. We emphasize that this phenomenon can be realized only when the physical Hamiltonian and the dissipation operators obey the symmetry preserving conditions we obtained in this work. So far, no general framework has been established to measure the density matrix topology. However, physical observables of density matrix topology have been proposed for specific cases, such as the ensemble geometric phase for AIII class in one dimension \cite{top_dens3}. This can be applied to our first example. How to experimentally observe such transitions in more general situations is still an open question. To this end, we show signatures in density matrix eigenvalues at the transition. Inspired by this spectrum property, we point out that the phase transition can be measured through the modified R\'enyi entropy. Moreover, our discussion so far is limited to Gaussian states and Lindbladians with quadratic Hamiltonian and linear dissipation operators. It will be interesting to study more general situations where the modular Hamiltonian hosts interacting symmetry protected topological phases. We leave these exciting issues for future studies.

\section{Acknowledgements}

 We thank Ying-Fei Gu, Zhong Wang and Tian-Shu Deng for helpful discussion. This work is supported by Innovation Program for Quantum Science and Technology 2021ZD0302005, the Beijing Outstanding Young Scholar Program and the XPLORER Prize. F.Y. is supported by Chinese International Postdoctoral Exchange Fellowship Program (Talent-introduction Program) and Shuimu Tsinghua Scholar Program at Tsinghua University.

\bibliographystyle{unsrt}

\end{document}